\documentclass[aps,pra,showpacs,superscriptaddress,twocolumn]{revtex4-1}

\usepackage{mathrsfs}
\usepackage{amsfonts}
\usepackage{amssymb}
\usepackage{amsmath}
\usepackage{graphicx}
\usepackage{dsfont}
\usepackage[usenames,dvipsnames]{color}
\usepackage[colorlinks=true,citecolor=blue,linkcolor=magenta]{hyperref}
\usepackage{ulem}

\renewcommand{\Im}{\mathrm{Im}}

\newcommand{\abs}[1]{| #1 |}

\newcommand{\ket}[1]{\vert{ #1 }\rangle}
\newcommand{\bra}[1]{\langle{ #1 }\vert}
\newcommand{\ketbra}[2]{\vert #1 \rangle \langle #2 \vert}
\newcommand{\braket}[2]{\langle #1 \vert #2 \rangle}

\newcommand{\V}[1]{\mathbf{#1}}
\newcommand{\VO}[1]{\mathbf{\hat{#1}}}
\newcommand{\mum}{\text{ }\mu\text{m}}
\newcommand{\mus}{\text{ }\mu\text{s}}
\newcommand{\pN}{\text{ pN}}
\newcommand{\force}{\text{ hGHz}/\mu\text{m}}

\newcommand{\MHz}{\text{ MHz}}
\newcommand{\kHz}{\text{ kHz}}
\newcommand{\vdW}{\text{ GHz}\mu\text{m}^6}

\begin{document}

\title{Spatial imaging of the movement of bound atoms to reveal the Rydberg molecular bond via electromagnetically induced transparency}

\author{Mingxia Huo}

\affiliation{Centre for Quantum Technologies, National University of Singapore, 3 Science Drive 2, 117543, Singapore}
\affiliation{Clarendon Laboratory, University of Oxford, Parks Road, Oxford OX1 3PU, United Kingdom}

\begin{abstract}
We propose an approach to detect individual Rydberg molecules with each molecule consisting of two atoms in different Rydberg states. The scheme exploits the movement of atoms in the presence of an external force that exerts only on atoms in one Rydberg state. Since the movement of atoms in the other Rydberg state depends on whether they are bound with atoms directly driven by the applied force, bound atoms can be distinguished from unbound atoms. By utilizing electromagnetically induced transparency, it is possible to non-destructively image the positions of molecules. The scheme is sensitive to a weak force, which is suited to optically detect spatial positions and bond structure of Rydberg molecules. 
\end{abstract}


\maketitle

Single-molecule force spectroscopy techniques have been utilized to measure the forces and motions of individual molecules, which have provided a potential to reveal their fundamental properties~\cite{weiss:99,gross:09,gross:10,ulmanis:12,oteyza:13,neuman:08,walter:08}. The forces associated with the measurements are usually $10^{-3} \sim 10^{4} \pN$~\cite{neuman:08}. Rydberg atoms are atoms where at least one electron is in a highly excited state \cite{gallagher:94}. Rydberg molecules can be formed due to the interaction of the Rydberg electron with ground-state atoms~\cite{greene:00,bendkowsky:09,liu:09,bendkowsky:10,li:11,krupp:14} or the long-range interaction between two Rydberg atoms~\cite{boisseau:02,schwettmann:06,schwettmann:07,overstreet:09,samboy:11,samboy:11b,kiffner:12,tallant:12,kiffner:13,kiffner:13b,kiffner:14,petrosyan:14}. The Rydberg molecular bonds are broken apart in order to detect them through ionization, which is the most common technique~\cite{bendkowsky:09,overstreet:09}. On the other hand, spatial distributions of Rydberg atoms have been resolved non-destructively via Rydberg electromagnetically induced transparency (EIT)~\cite{olmos:11,gunter:12,gunter:13}. By utilizing the non-destructive measurement of Rydberg atoms, we present a theoretical proposal to optically image spatial positions and access bond strengths of Rydberg molecules. 

We consider a typical Rydberg macrodimer in a stable energy minimum, surrounded by a region of attraction extending to some finite distance beyond which the potential becomes vanishing or repulsive~\cite{kiffner:14}. This bound potential results from avoided crossings caused by the long-range dipole-dipole interaction between two atoms in different Rydberg states~\cite{kiffner:14}. When the interatomic distance is larger than the region of attraction, the bond can be considered broken. A moderate force dragging one of two atoms in the molecule makes the other atom move due to their bond. In the presence of the applied dragging force, molecules and free atoms move different distances, which are observable by performing measurement on surrounding probe atoms. Our scheme only requires performing the imaging for two times: the first one is performed after molecule creation, which records the original positions of molecules; the second one is performed after dragging the molecules, which records the final positions of molecules. Our analysis shows that the timescale including the molecule creation, the imaging of surrounding  atoms, and the evolution under the applied dragging force should be comparable to the lifetime of Rydberg molecules. 

\begin{figure}[tph]
\includegraphics[width=8.5cm]{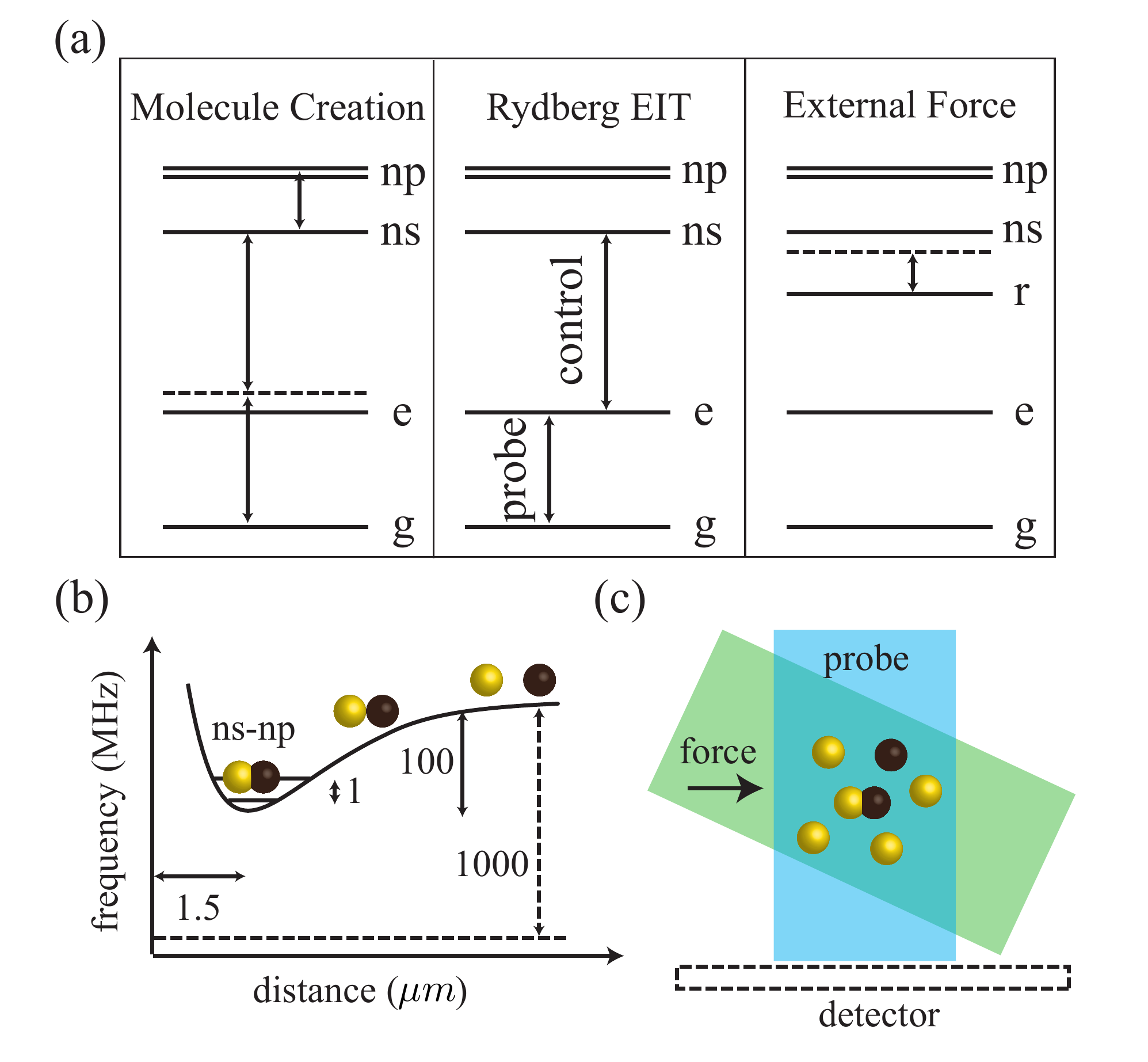}
\caption{\label{fig1}
(Color online) 
(a) Energy-level structures under consideration. $g$ and $e$ label the ground and excited states. $r$, $ns$, and $np$ represent Rydberg states. From left to right: molecules are created via two-photon transition followed by microwave excitations; a Rydberg EIT is available in the presence of probe and control lasers; a far off-resonance laser is introduced to move $ns$ atoms. (b) Diatomic molecular potential. A dissociation process is a competition between the potential force and the external force applied to $ns$ atoms. The vibrational frequency, the molecular potential depth, and the fine structure splitting are approximately on the order of $1$, $100$, and $1000$ $\mathrm{MHz}$, respectively. The potential minimum corresponds to an interatomic distance of $1.5$ $\mathrm{\mu m}$. (c) The applied force and the EIT probe beam employed for detection. 
}
\end{figure}

The Rydberg molecule under consideration consists of two atoms in different Rydberg states $ns$ and $np$~\cite{kiffner:14}. Here we label the atomic energy level as $nl_j$. For Rydberg atoms, as shown in Fig. \ref{fig1}(a), the principal quantum number $n\gg 1$. $l=s,p$ labels the orbital angular momentum of the excited valence electron. $j$ is the total angular momentum. The $ns$ level is separated from the $np_{3/2}$ level by energy $\hbar \omega_0$, while the $np_{1/2}$ level is separated from the $np_{3/2}$ level by fine structure splitting $\hbar \abs{\Delta}$. Here for Rb atoms, $\omega_{0}$ is much larger than $\Delta$~\cite{mourachko:03}. We propose to start with a dilute gas of cold Rb atoms prepared in a ground state $ 5s$. The atoms are excited to the Rydberg state $ns$ via a two-photon transition. This is followed by microwave radiations to excite pairs of $ns$ atoms to $ns-np$ molecules~\cite{kiffner:14}. The molecules are observable by detecting probe absorptions in an EIT transition via direct detection~\cite{gunter:13} or a fluorescence-detection scheme proposed in this work. Here when probe atoms are in a region surrounding the molecule, an energy shift of the Rydberg EIT level $ns$ of probe atoms allows absorption of EIT probe field, which gives a spot surrounding each molecule in the image. The unbound $ns$ and $np$ atoms also cause level shift of probe atoms. To distinguish molecules from unbound $ns$ and $np$ atoms, a detuned laser beam with an inhomogeneous intensity is switched on to move $ns$ atoms. After the evolution, the measurement of EIT off-resonance of probe atoms gives shifted spots, double-distance shifted spots, and unchanged spots in the images. They correspond to $ns-np$ molecules, $ns$ atoms, and $np$ atoms, respectively. The double-distance shift of $ns$ atoms is because the molecule mass is twice of the single-atom mass. We show that the observable rupture force of Rydberg molecular bond is on the order of $10^{-7} \pN$. 

The Hamiltonian describing two Rydberg atoms is given by $H_{\text{2atoms}} = (\VO{p}_1^2 + \VO{p}_2^2)/(2m) + H_0(\VO{R}_1,\VO{R}_2)$, where $\VO{p}_i$ is the momentum of the $i$-th atom and $m$ is the mass of a single atom. The internal Hamiltonian of two Rydberg atoms reads $H_0(\VO{R}_1,\VO{R}_2) = H_1+ H_2 + V_{1,2}(\VO{R}_1,\VO{R}_2)$, where $H_{i}$ is the internal Hamiltonian for the $i$-th atom, and $V_{1,2}(\VO{R}_1,\VO{R}_2)$ gives the dipole-dipole interaction. The explicit internal Hamiltonian form is given in Appendix A.

In the $nsnp$ subspace consisting of two atoms, with one atom in $ns$ state and the other atom in $np$ state, the molecular potential is revealed in a calculation of the internal eigenstates using the Born-Oppenheimer approximation~\cite{kiffner:12,kiffner:13,kiffner:14}, which has been verified via semiclassical calculations showing that the atomic motion in the potential well remains adiabatic~\cite{kiffner:13}. The reason is that the eigenstate corresponding to the potential well, i.e.~bound state, is energetically well separated from other eigenstates, and the velocity of atoms in the well is sufficiently small \cite{kiffner:13}. Other two-atom states cause a negligible van der Waals shift when $\abs{\V{R}} \geq r_{0}$, since their energy separation from the $nsnp$ manifold is large compared to $\hbar \vert \Delta\vert$. Here $\V{R}$ is the interatomic distance and $r_{0}$ is a characteristic length as defined in Appendix A. In Fig. \ref{fig1}(b), we schematically display a potential well with a minimum near $R_{p}\simeq 1.7r_{0}$. The well depth is on the order of $100$ $\mathrm{MHz}$ for $n=40$. An electric field can induce a Stark shift $\delta$ with $\vert \delta\vert\ll \vert \Delta\vert$, which allows one to align the molecule in the direction of the electric field \cite{kiffner:14}. 

We assume to start with a dilute gas of Rb ground atoms and expect to obtain Rydberg molecules with a certain successful probability~\cite{overstreet:09}. The resulting populations of Rydberg atoms and molecules are small but should allow experiments in a realistic time~\cite{overstreet:09}. In a short distance, the transition from one Rydberg excitation to two Rydberg excitations is shifted by the dipole-dipole interaction between two Rydberg atoms, which forms a band of energies. Because the interatomic distance in the molecule is known, i.e.~the blockade energy is known, one can utilize a laser with a different frequency to excite the second Rydberg atom. Actually, because the $ns-ns$ state is not the ultimate state, one can also directly excite the second atom to the $np$ state by utilizing the microwave, whose frequency is determined by taking into account the blockade effect, i.e.~given by the splitting of the $ns$ and $np$ states, the blockade energy due to the interaction between two $ns$ atoms, and the binding energy of the molecule due to the interaction between $ns$ and $np$ atoms. At a temperature of $300$ $\mathrm{\mu K}$, we estimate a decay rate of approximately $25$ $\mathrm{kHz}$ for $^{85}$Rb with $n=40$. The vibrational frequencies of the molecular motion are $\simeq 2\pi \times 0.4$ $\mathrm{MHz}$, and the rotational frequencies are $\simeq 2\pi \times 0.1$ $\mathrm{kHz}$~\cite{kiffner:14}. The equilibrium distance is given by $R_{p}=1.55$ $\mathrm{\mu m}$. A necessary condition for the detection of Rydberg molecules is that their kinetic energy is smaller than the depth of the potential well, which corresponds to a temperature of $\sim 4.8$ $\mathrm{mK}$ ($100$ $\mathrm{MHz}$) for $n=40$ \cite{kiffner:13}. 


We consider an EIT detection scheme, where the probe beam $\Omega_{p}$ and the control beam $\Omega_{c}$ are resonant to the transitions of $g \leftrightarrow e$ and $e \leftrightarrow ns$, respectively [Fig.~\ref{fig1}~(a) and Appendix B. Here $g$ and $e$ denote the ground and excited states, respectively. We assume the state $e$ decays into $g$ and another ground state $g'$ with a same decay rate. The state $g'$ is decoupled from the EIT lasers, which is detectable with fluorescence. When a probe atom is far from the region around one impurity, i.e.~a Rydberg atom or a molecule, the imaginary part of susceptibility $\Im[\chi]$ (Appendix B) and the $g'$-state population of the probe atom are zero. When the probe atom is in the vicinity of the impurity, the $ns$ level of the probe atom is shifted due to the dipole-dipole interaction, leading to a finite $\Im[\chi]$ and $g'$-state population. 

\begin{figure}[t!]
\includegraphics[width=8.5cm]{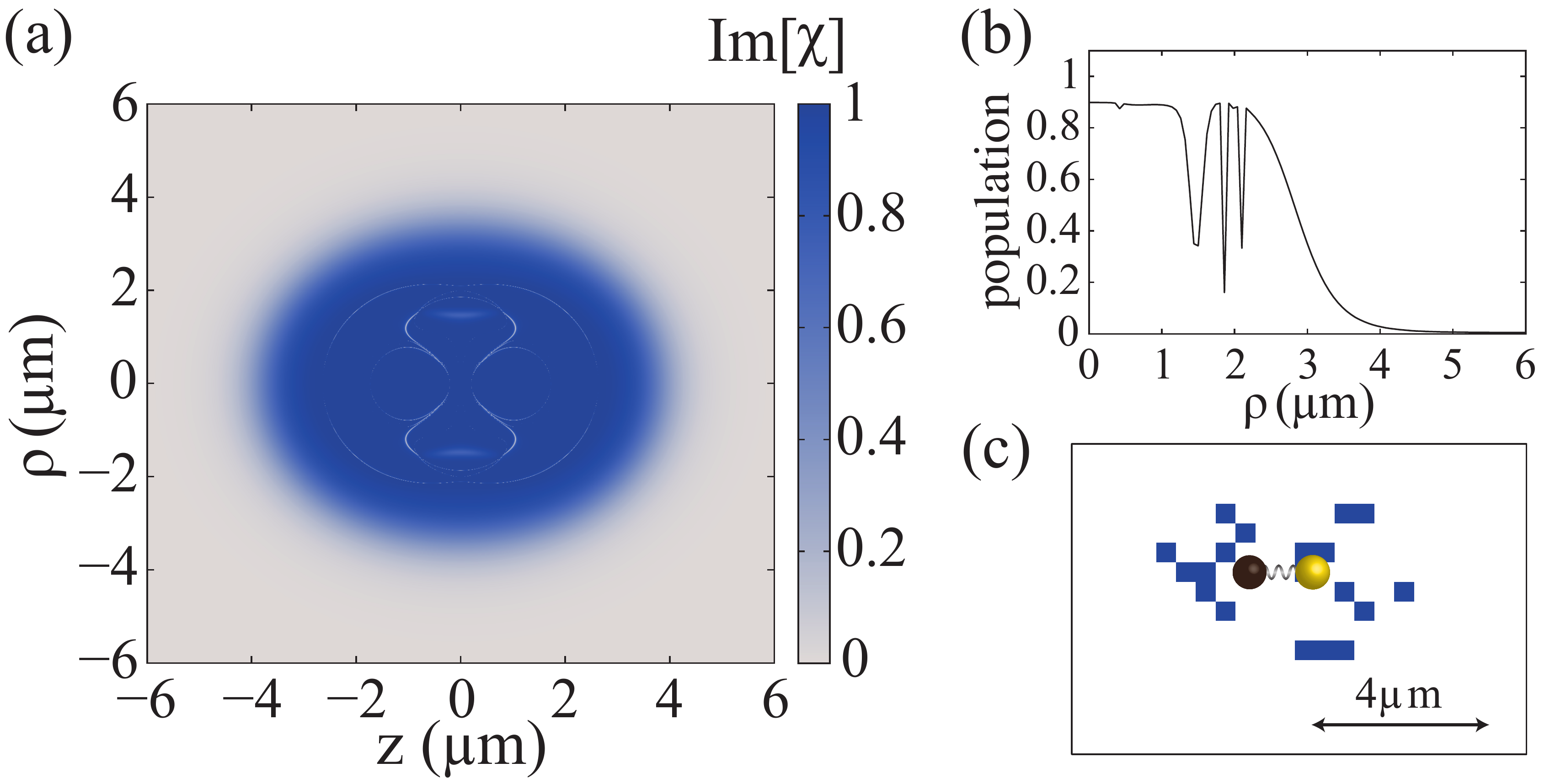}
\caption{\label{fig2}
(Color online)
(a) The imaginary part of susceptibility and (b) the $g'$-state population as functions of the probe-atom position. Two bound atoms are located at $\rho=0,z=\pm 0.8\mum$. $\Im[\chi]$ approaches to $1$ when $z,\rho <3\mum$, corresponding to a finite probe absorption. The $ g'$-state population has a similar behaviour. Out of this region, $\Im[\chi]$ and the population drop rapidly. (c) Simulated image including atomic density fluctuations at each spatial position. The position of molecule is resolved by surrounding fluorescence blue spots.
}
\end{figure}

We consider one molecule with two atoms positioned at $\rho=0,z=\pm R_{p}/2$, and one probe atom with varying spatial positions, as shown in Fig.~\ref{fig2}(a). $\Im[\chi]$ is zero when the probe atom is far from the region around the molecule. In the vicinity of the molecule within a radius of $R_{c}\sim 3\mum$, $\Im[\chi]$ becomes finite. We would like to note that, several lines with $\Im[\chi]$ significantly smaller than $1$ appear in this region due to the existence of degenerate states to the state in which the probe atom in the $ns$ state is far from the molecule. However, these lines are very thin. In the calculation, we assume $\Omega_{p} = 2\pi \times 1 \MHz$ and $\Omega_{c} = 2\pi \times 10 \MHz$. The decay rates from $e$ and $ns$ states are $\Gamma_{p} = 2\pi \times 6 \MHz$ and $\Gamma_{c} = 2\pi \times 25 \kHz$, respectively. 

By numerically solving the master equation, the $g'$-state population after switching on EIT lasers for $2\mus$ is plotted in Fig.~\ref{fig2}(b). Since the distribution of $ g'$-state population in $z-\rho$ plane has a similar shape as that of $\Im[\chi]$ in (a), we plot the population for $z=0$ in (b) as an illustration. When the population of $ns$ atoms is small, i.e.~$\rho_{\mathrm{2D}}R_{c}^2\abs{\Omega_{p}/\Omega_{c}}^2 \ll 1$, with $\rho_{\mathrm{2D}}$ the density of atoms in 2D, the results for single probe atom can be applied to an ensemble of probe atoms. By including the atomic density fluctuation, a spatially resolved fluorescence image is simulated in Fig.~\ref{fig2}(c), where we assume $\rho_{\mathrm{2D}} = 1\mum^{-2}$, and each pixel has a region of $(0.5\mum)^2$. As illustrated in Fig.~\ref{fig2}(c), the position of molecule is surrounded by blue spots, with each spot representing one $g'$ atom. One molecule is surrounded by approximately $20$ $g'$ atoms on average.

To distinguish molecules from Rydberg atoms, we assume to drag the molecules for an resolvable distance of $9$ $\mu m$ by applying an energy slope of e.g.~$0.08 \force$. This slope can be realized with a detuning of $20$ $\mathrm{GHz}$ and a Rabi frequency increasing linearly from $0$ to $8$ $\mathrm{GHz}$ over a distance of $10$ $\mathrm{\mu m}$. With these parameters, the depopulation of the Rydberg state $ns$ occurs with a probability of $4\%$. If an inhomogeneous laser is driving a transition between a low-energy state and the $ns$ state, the required Rabi frequency of several $\mathrm{GHz}$ is challenging. However, if the laser is driving a transition between two Rydberg states, the laser power can be significantly reduced, because the dipole matrix elements can be much larger. The energy of $np$ state is also shifted, however, the magnitude of the shift is much smaller. Specifically, $\vert \Delta\vert$ is not much larger than the detuning of the transition involving the $ns$ state, e.g. a transition of $ns\leftrightarrow n'p$. Fortunately, due to the selection rule, the detunings of the transitions involving the $np$ state are determined by the splittings in the $n'$ level: there are only transitions of $np\leftrightarrow n's$ and $np\leftrightarrow n'd$. The detunings for these transitions are determined by the energy differences of the $n's$, $n'd$, and $n'p$ states. When $n'$ is significantly smaller than $n$, the splittings in $n'$ levels are much larger than that in $n$ levels, i.e.~the detunings for the transitions involving the $np$ state are much larger than that involving the $ns$ state.  

The center of mass motion of two atoms is decoupled from the internal state. For the relative motion, the frequencies of vibrations and rotations are $< 1$ $\mathrm{MHz}$. They are much smaller than the minimum separation between the bound state and other eigenstates. In the presence of the applied force, the dissociation process is a competition between the dragging and potential forces. The internal state evolves adiabatically, where the system stays in an eigenstate adiabatically connected to the initial bound state.  We label the molecular bound state as $\psi(\V{R}_1,\V{R}_2)$. Under the Born-Oppenheimer approximation, we describe the motion of two atoms with a classical Hamiltonian of $\bar{H}_{\text{2atoms}}(\V{R}_1,\V{R}_2) = (\V{p}_1^2 + \V{p}_2^2)/(2m) + \bar{H}_0(\V{R}_1,\V{R}_2)$, where $\bar{H}_0(\V{R}_1,\V{R}_2) = \bra{\psi(\V{R}_1,\V{R}_2)} [H_0(\V{R}_1,\V{R}_2)+U_1+U_2] \ket{\psi(\V{R}_1,\V{R}_2)}$. The applied force on the $i$-th atom is denoted by $U_i = u(\V{R}_i)\sum_{j_z}\vert{ns,j_z}\rangle_{i} \langle{ns,j_z}\vert_i$ with $u(\V{R}_i)$ the energy shift of $ns$ state of the $i$-th atom. Here $j_z$ is the projection of the total angular momentum onto the $z$ axis. In Hamiltonian mechanics, the time evolution of the classical system is described by the Hamilton's equations of $d\V{p}_i / dt = - \nabla_{\V{R}_i} \bar{H}_0(\V{R}_1,\V{R}_2)$ and $d\V{R}_i / dt = \V{p}_i / m$.

\begin{figure}[t!]
\includegraphics[width=8.5cm]{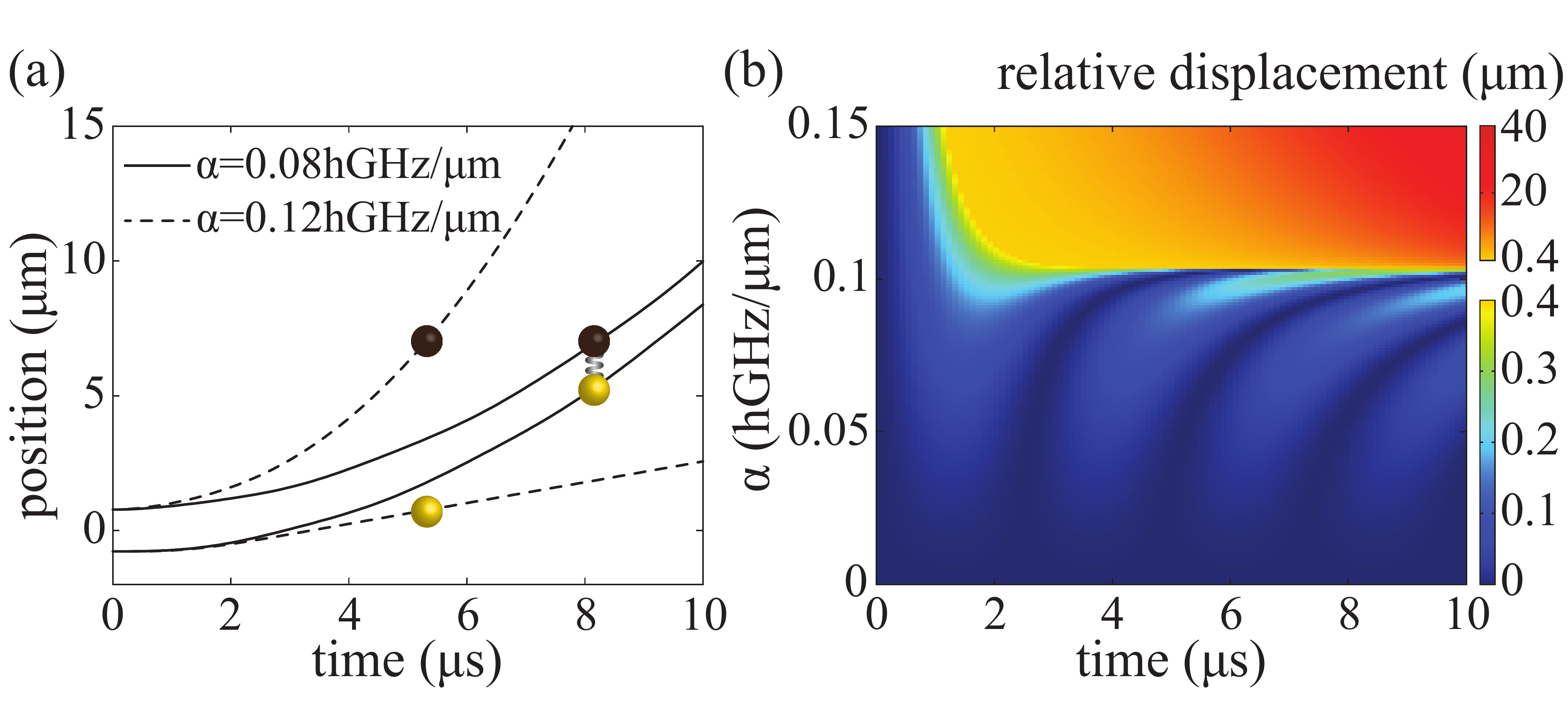}
\caption{\label{fig3}
(Color online)
(a) Atomic position versus time. In the presence of a small dragging force with $\alpha = 0.08 \force$ (solid line), two atoms move together by $9\mum$ in $10\mus$. When a larger dragging force with $\alpha = 0.12 \force$ is applied (dashed line), the molecular bond breaks apart. (b) Relative atomic displacement versus time and $\alpha$. The oscillatory behaviour along the time axis for small $\alpha$ indicates an unbroken bond between two atoms. The bond breaks apart when $\alpha$ exceeds $0.1 \force$.
}
\end{figure}

The dragging force is applied by a linear energy shift $u = \alpha z$. We consider the case that the molecule is aligned along the direction of the applied force because in this case, the molecular bond breaks apart with the minimum force strength. The dynamics for the molecule aligned along other directions is similar but the molecular bond allows for a larger force strength. With two atoms initialized at $x=y=0$, $z=\pm R_{p}/2$, their positions as a function of time is given in Fig.~\ref{fig3}(a). The solid line corresponds to the case of a small energy shift with $\alpha = 0.08 \force \simeq 5\times 10^{-7} \pN$, where two atoms move together by $9\mum$ in $10\mus$. For a larger force with $\alpha =0.12\force$ (dashed line), two atoms begin to drift away from each other, indicating that the molecular bond breaks apart. As shown in Fig.~\ref{fig3}(b), an oscillatory behaviour of relative displacement of two atoms occurs for small $\alpha$, where two atoms are bound together. When $\alpha$ exceeds $0.1\force$, the relative displacement becomes very large without oscillation, indicating a broken bond. We would like to note that, as shown in Fig. \ref{fig3}(b), the frequency of the relative motion in the presence of the applied force is $< 1$ $\mathrm{MHz}$. We have calculated the overlap between $\psi(\V{R}_1,\V{R}_2)$ after switching on an applied force with  $\alpha=0.08 \force$ and the initial bound state, where the overlap is higher than $98\%$. Therefore, the internal state stays in an eigenstate adiabatically connected to the initial bound eigenstate with a high probability. Once the molecule is projected into other states, two atoms separate up, which should be detectable.

The interaction between probe atoms is the van der Waals interaction $V \simeq - \hbar C_{6} / R^{6}$, with $C_{6} = 2\pi \times 1 \vdW$ for $40s$ level considered here~\cite{singer:05}, giving a critical radius of $R'_{c} = (2C_{6}\Gamma_{p}/\Omega_{c}^{2})^{1/6} \simeq 2\mum$. The radius $R'_{c}$ can be shortened by choosing an EIT level of $n's$ with $1\ll n'<40$ for probe atoms, where the interaction coefficient $C_{6}$ decreases as its leading term is proportional to $n^{11}$~\cite{singer:05}. A large signal-to-noise ratio imposes a constraint of $\rho_{\mathrm{2D}}\lesssim \Omega_{c}^{2}/ (\pi R'^{2}_{c} \Omega^{2}_{p})$, above which the contrast of the image decreases \cite{gunter:12}. In order to implement the imaging scheme, we propose to use a 2D gas with a density of $\rho_{\mathrm{2D}}=1$ $\mathrm{\mu m}^{-2}$, $R'_{c}=2$ $\mathrm{\mu m}$, and $\Omega_{c}^{2}/\Omega_{p}^{2}=100$, in which the distance between atoms is short enough for creating molecules and the constraint for a large signal-to-noise is satisfied. In Ref.~\cite{gunter:13}, the exposure times for imaging Rydberg atoms are $2-20$ $\mathrm{\mu s}$, and it is expected to ultimately follow the evolution of individual atoms in real time. The amplification factor of one impurity in Ref.~\cite{gunter:13}, which is related to the optical depth, is approximately equal to the number of probe atoms in the $g'$ state surrounding one molecule in our proposal. 

In conclusion, we have theoretically proposed a scheme to optically detect Rydberg molecules. With an applied dragging force, Rydberg molecules are distinguished from unbound atoms with distinct features in images of probe absorptions in an EIT transition via direct detection [38] or a fluorescence-detection scheme proposed in this work.  We have shown that a Rydberg molecular bond remains unbroken in the presence of an applied force when its strength is within a range. The proposed measurement is performed on surrounding atoms, which allows for non-destructive and all-optical investigations of Rydberg molecules. The unbroken molecular bond for spatial detection in principle allows for access to dynamics of individual Rydberg molecules. The principle can be applied to other kinds of Rydberg molecules. It has potential for revealing Rydberg molecular bond properties, dynamics in real time, spatial correlations of Rydberg molecules, as well as quantum simulation based on Rydberg molecules.

\begin{acknowledgements}
The author is grateful to Dieter Jaksch for very helpful discussions as well as valuable suggestions. The author acknowledges financial support from National Research Foundation and Ministry of Education, Singapore.
\end{acknowledgements}

\appendix

\section{Hamiltonian and molecular potential}

The Hamiltonian of the $i$-th Rydberg atom is given by
\begin{eqnarray}
H_{i} &=& \hbar \omega_{0} \sum_{j_z=-3/2}^{3/2}\ketbra{np_{3/2},j_z}{np_{3/2},j_z}_{i} \notag \\
&&+ \hbar (\omega_{0}+\Delta) \sum_{j_{z}=-1/2}^{1/2}\ketbra{np_{1/2},j_z}{np_{1/2},j_z}_{i},
\end{eqnarray}
where $\omega_{0}$ is the resonance frequency of the $ns \leftrightarrow np_{3/2}$ transition, and $\hbar \abs{\Delta}$ is the energy spacing between the $np_{1/2}$ and $np_{3/2}$ states. The dipole-dipole interaction~\cite{tannoudji:api} between atoms $i$ and $j$ located at
positions $\V{R}_{i}$ and $\V{R}_{j}$ is defined as
\begin{equation}
V_{ij} = \frac{1}{4\pi\varepsilon_0 R^3}
[ \VO{d}^{(i)} \cdot \VO{d}^{(j)}
- 3 (\VO{d}^{(i)} \cdot \vec{\V{R}}) (\VO{d}^{(j)} \cdot \vec{\V{R}}) ].
\end{equation}
Here $\epsilon_{0}$ is the dielectric constant. $\VO{d}^{(i)}$ is the electric dipole-moment operator of the $i$-th atom. $\V{R} = \V{R}_{i} - \V{R}_{j}$, $R=\vert \V{R}\vert$, and $\vec{\V{R}} = \V{R} / R$ is the corresponding unit vector. The matrix elements of the electric-dipole-moment operator $\VO{d}$ of an individual atom are evaluated via the Wigner-Eckert theorem~\cite{walker:08,edmonds:amq}. We define the reduced dipole matrix element as $D = e\langle np \vert r\vert ns\rangle / \sqrt{3}$. For alkali-metal atoms with $n\geq 40$, we have $\langle np\vert r\vert ns\rangle \simeq n^{2}a_{0}$, where $a_{0}$ is the Bohr radius~\cite{walker:08}. The characteristic strength of the dipole-dipole interaction is given by $\hbar \Omega = \vert D\vert ^{2} / (4\pi \epsilon_{0}R^{3})$. The characteristic length scale $r_{0}$ at which bound states occur is obtained by equating $\Omega$ with $\vert \Delta\vert$. This gives $r_{0}=[ \vert D\vert ^{2} / (4\pi \epsilon_{0} \hbar \vert \Delta \vert) ]^{1/3}$. For $^{85}$Rb with $n=40$, the splitting is $\Delta\simeq 2\pi \times 1$ $\mathrm{GHz}$, which yields $r_{0}\simeq 1$ $\mathrm{\mu m}$ \cite{kiffner:14}.

\section{Susceptibility}

The Hamiltonian describing the interaction between a probe atom and EIT lasers is
\begin{eqnarray}
H_{\text{EIT}} &=& \hbar \sum_{j_z=-1/2}^{1/2}[\Delta_{p} \ketbra{e,j_z}{e,j_z} \notag \\
&&+(\Delta_{p}+\Delta_{c}) \ketbra{ns,j_z}{ns,j_z} \notag \\
&& + \frac{ \Omega_{p}}{2} ( \ketbra{e,j_z}{g,j_z}+\text{H.c.}) \notag \\
&&+ \frac{\Omega_{c}}{2} (\ketbra{ns,j_z}{e,j_z} + \text{H.c.} )],
\end{eqnarray}
where $g$ and $e$ represent the ground and excited states. The probe $\Omega_{p}$ and control $\Omega_{c}$ lasers drive the $g \leftrightarrow e$ and $e \leftrightarrow ns$ transitions with detunings $\Delta_{p}$ and $\Delta_{c}$, respectively.  $\Gamma_{p}$ and $\Gamma_{c}$ are the spontaneous decay rates of the $e$ and $ns$ states. 
 
In the limit of weak probe $\Omega_{p} \ll \Omega_{c}$ and small excitations to the $e$ and $ns$ states, the susceptibility corresponding to the probe transition is~\cite{badger:01,sevincli:11,gunter:12}
\begin{eqnarray}
\chi = \frac{i\Gamma_p}
{ \Gamma_p - i\Delta_p
+ \sum_k \Omega_c^2 F_{k}^2 (\Gamma_c - i\Delta_k)^{-1} },
\label{chi}
\end{eqnarray}
where the sum over $k$ accounts for the effect of multiple energy levels $\vert \psi_{k}\rangle$ of three-Rydberg-atom states~\cite{badger:01}. We label the $k$-th eigenenergy and eigenstate of three Rydberg atoms as $E_{k}$ and $\vert \psi_{k}\rangle$, respectively. $F_{k}^2=\sum_{j_z=-1/2}^{1/2}\abs{\braket{ns,j_z,\text{m}}{\psi_k}}^2$ with $\ket{ns,j_z,\text{m}}$ denotes a state in which the probe atom is in the $\ket{ns,j_z}$ state and the other two atoms are in the molecular state. $E_{0}$ is the energy of the $\ket{ns,j_z,\text{m}}$ state when the probe atom is far away from the molecule. The detuning $\Delta_{k}=\Delta_{p}+\Delta_{c}+E_{k}-E_{0}$.

\end{document}